\documentstyle[12pt]{article}
\def\mathbf{\vec}

\def\ca{\c{c}\~{a}}

\begin{document}

\centerline {\LARGE One-loop fermion determinant with}
\vspace{.3cm}
\centerline {\LARGE explicit chiral symmetry breaking}
\vspace{1cm}
\centerline {\large Alexander A. Osipov\footnote{On leave from the 
            Laboratory of Nuclear Problems, JINR, Dubna, Russia}, 
            Brigitte Hiller}
\vspace{.5cm}
\centerline {\it Centro de F\'{\i}sica Te\'{o}rica, Departamento de
             F\'{\i}sica}
\centerline {\it da Universidade de Coimbra, 3004-516 Coimbra, Portugal}
\vspace{1cm}

\begin{abstract}
We use a proper-time regularization to define the one-loop fermion determinant
for the case in which {\it explicit} chiral symmetry breaking takes place. We 
show how to obtain the polynomial by which the standard definition of $\ln\det 
D$ needs to be modified in order to arrive at the fermion determinant whose 
transformation properties are consistent with the general symmetry requirements
of the basic Lagrangian. As an example it is shown how the fundamental 
symmetries and the explicit chiral symmetry breaking pattern associated with 
the ENJL model are preserved in a consistent way.
\end{abstract}


It is known \cite{Ball:1989} how to derive an exact non-perturbative
representation for the chiral fermion determinant which (modulo anomalies)
is manifestly chiral gauge covariant. In particular this technique has
been widely used in the literature \cite{Ebert:1986,Bijnens:1993} to derive
the low-energy effective action of an extended Nambu -- Jona-Lasinio (ENJL)
model with the explicit chiral symmetry breaking term in the Lagrangian. 
The central object in the calculation of the effective action is the quantity
$\ln\det D$, where the differential Dirac operator $D$ depends on collective
meson fields which have well defined transformation laws with respect to the 
action of the chiral group. If one neglects the current quark mass term in
$D$ the combination $D^\dagger D$ transforms covariantly. This fact ensures
that the definition of $\ln\det D$ in terms of a proper time integral
\begin{equation}
\label{logdet}
   \ln\det D=-\frac{1}{2}\int^\infty_0\frac{dT}{T}\rho (T,\Lambda^2)\mbox{Tr}
              \left(e^{-TD^\dagger D}\right)
\end{equation}
cannot destroy the symmetry properties of the basic Lagrangian.

In this note we observe that in the presence of the explicit chiral symmetry 
breaking term this is not any longer true. The naive definition of $\ln\det D$ 
in terms of a proper time integral (\ref{logdet}) modifies the chiral symmetry 
breaking pattern of the original quark Lagrangian and needs to be corrected in 
order to lead to the fermion determinant whose transformation properties 
exactly comply with the symmetry content of the basic Lagrangian. The necessary
modification can be done by adding a polynomial in the collective fields and 
their derivatives to the right hand side of eq.(\ref{logdet}) in  full analogy
with Gasser and Leutwyler's calculations in \cite{Gasser:1984}. This polynomial
must be chosen in such a manner that the real part of the effective Lagrangian 
for the bosonized ENJL model ${\cal L}_{\mbox{eff}}$ will have the same 
transformation laws as the basic quark Lagrangian ${\cal L}$. This requirement 
allows one to test if the explicit chiral symmetry breaking effect has been 
correctly calculated and we find that the expressions given in the literature 
are incorrect.

Consider the effective quark Lagrangian of strong interactions which is
invariant under a global colour $SU(N_c)$ symmetry
\begin{eqnarray}
\label{enjl}
  {\cal L}&=&\bar{q}(i\gamma^\mu\partial_\mu -\hat{m})q
            +\frac{G_S}{2}[(\bar{q}q)^2+(\bar{q}i\gamma_5\tau_i q)^2]
            \nonumber\\
          &-&\frac{G_V}{2}[(\bar{q}\gamma^\mu\tau_i q)^2
            +(\bar{q}\gamma^\mu\gamma_5\tau_i q)^2].
\end{eqnarray}
Here $q$ is a flavour doublet of Dirac spinors for quark fields $\bar q=(\bar 
u, \bar d)$. Summation over the colour indices is implicit. We use the standard
notation for the isospin Pauli matrices $\tau_i$. The current quark mass matrix
$\hat{m}=\mbox{diag}(m_u, m_d)$ is chosen in such a way that $m_u=m_d$. Without
this term the Lagrangian (\ref{enjl}) would be invariant under global chiral 
$SU(2)\times SU(2)$ symmetry.

The transformation law for the quark fields is the following
\begin{equation}
\label{quark}
   \delta q=i(\alpha +\gamma_5\beta )q, \quad
   \delta\bar{q}=-i\bar{q}(\alpha -\gamma_5\beta )
\end{equation}
where parameters of global infinitesimal chiral transformations are chosen as
$\alpha =\alpha_i\tau_i, \ \ \beta =\beta_i\tau_i$. Therefore our basic
Lagrangian ${\cal L}$ transforms according to the law
\begin{equation}
\label{sb}
   \delta {\cal L}=-2i\hat{m}(\bar{q}\gamma_5\beta q).
\end{equation}
It is clear that nothing must destroy this symmetry breaking requirement of the
model (we are not considering anomalies here).

Following the standard procedure we introduce colour singlet collective
bosonic fields in such a way that the action becomes bilinear in the quark
fields and the quark integration becomes trivial
\begin{eqnarray}
\label{gf1}
   Z&=&\int {\cal D}q{\cal D}\bar{q}{\cal D}s{\cal D}p_i{\cal D}V_\mu^i
      {\cal D}A_\mu^i
      \mbox{exp}\left\{i\int d^4x\left[{\cal L} \right.\right.\nonumber \\
    &-&\left.\left.\frac{1}{2G_S}(s^2+p_i^2)+\frac{1}{2G_V}
      (V_{\mu i}^2+A_{\mu i}^2)\right]\right\}.
\end{eqnarray}
We suppress external sources in the generating functional $Z$ and assume 
summation over repeated Lorentz $(\mu )$ and isospin $(i=1,2,3)$ indices. One 
has to require from the new collective variables that
\begin{equation}
\label{req}
   \delta (s^2+p_i^2)=0, \quad
   \delta (V_{\mu i}^2+A_{\mu i}^2)=0
\end{equation}
in order not to destroy the symmetry of the basic Lagrangian ${\cal L}$.
After replacement of variables
\begin{equation}
   s=\sigma -\hat{m}+G_S(\bar{q}q),
\end{equation}
\begin{equation}
   p_i=\pi_i -G_S(\bar{q}i\gamma_5\tau_iq),
\end{equation}
\begin{equation}
   V_{\mu}^i=v_\mu^i+G_V(\bar{q}\gamma_\mu\tau_iq),
\end{equation}
\begin{equation}
   A_{\mu}^i=a_\mu^i+G_V(\bar{q}\gamma_\mu\gamma_5\tau_iq),
\end{equation}
these requirements together with (\ref{quark}) lead to the transformation laws
for the new collective fields
\begin{equation}
\label{st}
   \delta\sigma =-\{\beta, \pi\}, \quad
   \delta\pi =i[\alpha, \pi ]+2(\sigma -\hat{m})\beta ,
\end{equation}
\begin{equation}
   \delta v_\mu =i[\alpha, v_\mu ]+i[\beta , a_\mu ], \quad
   \delta a_\mu =i[\alpha, a_\mu ]+i[\beta , v_\mu ].
\end{equation}
We have introduced the notation $\pi=\pi_i\tau_i$, $v_\mu=v_{\mu i}\tau_i$, 
$a_\mu=a_{\mu i}\tau_i$. Therefore the transformation law of the quark fields 
finally defines the transformation law of the bosonic fields.

The Lagrangian in the new variables has the form
\begin{eqnarray}
\label{lq}
  {\cal L}&=&\bar{q}[i\gamma^\mu\partial_\mu -\sigma +i\gamma_5\pi +
            \gamma^\mu (v_\mu +\gamma_5 a_\mu )]q \nonumber \\
          &-&\frac{(\sigma -\hat{m})^2+\pi_i^2}{2G_S}+
            \frac{v_{\mu i}^2+a_{\mu i}^2}{2G_V}.
\end{eqnarray}  
The subsequent integration over quark fields shows that the effective potential
has a non-trivial minimum and that spontaneous chiral symmetry breaking takes
place. Redefining the scalar field $\sigma\rightarrow\sigma +m$ we come finally
to the effective action
\begin{equation}
\label{seff}
   S_{\mbox{eff}}=-i\ln\det D-\int d^4x
   \left[\frac{(\sigma +m-\hat{m})^2+\pi_i^2}{2G_S}-
         \frac{v_{\mu i}^2+a_{\mu i}^2}{2G_V}\right]
\end{equation}
where the Dirac operator $D$ is equal to
\begin{equation}
   D=i\gamma^\mu\partial_\mu -m-\sigma +i\gamma_5\pi +
      \gamma^\mu (v_\mu +\gamma_5 a_\mu ).
\end{equation}
In this broken phase the transformation law of the pion field changes to 
\begin{equation}
   \delta\pi =i[\alpha, \pi ]+2(\sigma +m-\hat{m})\beta
\end{equation}
in full agreement with the variable replacement $\sigma\rightarrow\sigma +m$ 
for the scalar field in (\ref{st}).

The $\ln\det D$ is conveniently calculated using the heat kernel method or even
more directly in the way suggested in \cite{Ball:1989}. The result of these 
calculations on the basis of the formula (\ref{logdet}) is well known, see for 
example \cite{Ebert:1986}. We give it here using our notation and restricting 
to the second order heat coefficient,
\begin{eqnarray}
\label{leff}
   {\cal L}_{\mbox{eff}}
   &=&\frac{v_{\mu i}^2+a_{\mu i}^2}{2G_V}-\frac{1}{2G_S}
     [(\sigma +m-\hat{m})^2+\pi^2_i ]+\frac{N_cJ_0}{4\pi^2}(\sigma^2+2m\sigma
     +\pi^2_i) \nonumber \\
   &-&\frac{N_cJ_1}{8\pi^2}\left[\frac{1}{6}\mbox{tr}(v_{\mu\nu}^2
     +a_{\mu\nu}^2)-\frac{1}{2}\mbox{tr}\left((\nabla_\mu\pi )^2
     +(\nabla_\mu\sigma )^2\right)\right.\nonumber \\
   &+&\left.(\sigma^2+2m\sigma +\pi^2_i)^2\right]
\end{eqnarray}
where trace is to be taken in isospin space. Here we have used the notation
\begin{equation}
   v_{\mu\nu}=\partial_\mu v_\nu -\partial_\nu v_\mu -i[v_\mu ,v_\nu ]
             -i[a_\mu ,a_\nu ],
\end{equation}     
\begin{equation}
   a_{\mu\nu}=\partial_\mu a_\nu -\partial_\nu a_\mu -i[a_\mu ,v_\nu ]
             -i[v_\mu ,a_\nu ],
\end{equation}     
\begin{equation}
   \nabla_\mu\sigma =\partial_\mu\sigma -i[v_\mu ,\sigma ]+\{a_\mu ,\pi\},
\end{equation}    
\begin{equation}
   \nabla_\mu\pi =\partial_\mu\pi -i[v_\mu ,\pi ]-\{a_\mu ,\sigma +m\}.
\end{equation}
The functions $J_n$ represent the integrals which appear in the result of the
asymptotic expansion of the heat kernel
\begin{equation}
J_n=\int^\infty_0\frac{dT}{T^{2-n}}e^{-Tm^2}\rho (T,\Lambda^2),
         \quad n=0,1,2...
\end{equation}
We consider a class of regularization schemes (proper-time regularizations)
which can be incorporated in this expression through the kernel $\rho (T,
\Lambda^2)$. These regularizations allow to shift in loop momenta. A typical
example is the covariant Pauli-Villars cutoff \cite{Pauli:1949}
\begin{equation}
\label{cc}
      \rho (T, \Lambda^2)=1-(1+T\Lambda^2)e^{-T\Lambda^2}.
\end{equation}

The Lagrangian (\ref{leff}) obtained on the basis of formula (\ref{logdet}) 
does not fulfil the transformation law (\ref{sb}) and should be modified. 
Indeed, if one uses the classical equation of motion for the pion field, 
$\pi_i=iG_S\bar{q}\gamma_5\tau_i q$, see (\ref{lq}), one can rewrite
eq.(\ref{sb}) in terms of meson fields
\begin{equation}
\label{sbm}
   \delta {\cal L}=-\frac{2\hat{m}}{G_S}(\beta_i\pi_i ).
\end{equation}
Now it is a simple task to see that the Lagrangian (\ref{leff}) has a different
transformation law. We have calculated the corresponding polynomial
$P(\sigma ,\pi ,v_\mu ,a_\mu )$ which has to be included in the naive
definition (\ref{logdet}) to get the correct answer (\ref{sbm}). Let us 
note that $P$ is unique up to a chirally invariant polynomial. One can 
always choose $P$ in such a manner that the gap equation is not 
modified, i.e. using this chiral symmetry freedom to avoid from $P$ 
terms linear in $\sigma$. As a result the Lagrangian (\ref{leff}) gets an 
additional contribution
\begin{equation}
   {\cal L}_{\mbox{eff}}\rightarrow{\cal L}_{\mbox{eff}}
            +\Delta{\cal L}_{\mbox{eff}}
\end{equation}
where
\begin{eqnarray}
  \Delta{\cal L}_{\mbox{eff}}
  &=&-\frac{\hat{m}^2(\sigma^2+\pi_i^2 )}{2m(m-\hat{m})G_S}
    +\hat{m}\frac{N_cJ_1}{2\pi^2}[(2m-\hat{m})\sigma^2+\sigma
    (\sigma^2+\pi_i^2 )] \nonumber \\
  &-&\hat{m}\frac{N_cJ_1}{4\pi^2}\mbox{tr}\{(2m-\hat{m})a^2_\mu
    -a_\mu\partial_\mu\pi +ia_\mu [v_\mu ,\pi ]+2\sigma a^2_\mu\}.
\end{eqnarray}

Having established this counterterm one may look for kinetic and mass terms of
the composite meson fields and extract the physical meson masses by bringing 
the kinetic terms to the canonical form by means of field renormalizations. 
This field redefinitions are standard and we are not going to discuss them 
here, leaving this issue for a forthcoming more detailed and longer paper. Let 
us only point out that after redefinitions the symmetry breaking part takes 
the form \cite{Gas:1969}
\begin{equation}
   \delta{\cal L}_{\mbox{eff}}=-2m^2_\pi f_\pi (\beta_i\pi_i')
\end{equation}
which leads to the well known PCAC relation. Here $\pi_i'$ denotes the physical
pion field.

In conclusion we have analyzed in this work the effect of explicit chiral 
symmetry breaking in a bozonized version of $SU(2)$ ENJL model. We have shown 
that one cannot naively infer that the standard proper-time procedure 
formulated in terms of $D^\dagger D$ to evaluate the fermion determinant in 
the chiral symmetric case can be as well applied in the presence of explicit 
symmetry breaking terms. This naive procedure which is commonly used in the
literature is misleading. We have found that the most appropriate way to
trace the symmetry breaking pattern during the bosonization procedure is 
to use the definition of the one-loop fermion determinant in the form
suggested by Gasser and Leutwyler some years ago \cite{Gasser:1984}
extending it to be applicable to non-renormalizable models \cite{Osipov:2000}.
We have shown that in the case of explicit chiral symmetry breaking
it is necessary to modify the real part of $\ln\det D$ by the polynomial
$P(v,a,\sigma ,\pi )$ to get correctly the chiral invariant result
as well as the symmetry breaking part.


\section{Acknowledgements}

This work is supported by grants provided by Funda\ca o para a Ci\^encia e a
Tecnologia, PRAXIS/C/FIS/12247/1998, PESO/P/PRO/15127/1999 and NATO
"Outreach" Cooperation Program.

\baselineskip 12pt plus 2pt minus 2pt

\end{document}